\def\year{2019}
\documentclass[letterpaper]{article} 
\usepackage{aaai19}  
\usepackage{times}  
\usepackage{helvet}  
\usepackage{courier} 
\usepackage{url}  
\usepackage{graphicx}
\usepackage{amsmath}
\usepackage{multirow}

\begin{document}

\title{TransNFCM: Translation-Based Neural Fashion Compatibility Modeling}
\author{Xun Yang$^{\dag}$, Yunshan Ma$^{\dag}$, Lizi Liao$^{\dag}$, Meng Wang$^{\ddag}$, Tat-Seng Chua$^{\dag}$\\
$^{\dag}$School of Computing, National University of Singapore, Singapore\\
$^{\ddag}$School of Computing and Information Engineering, Hefei University of Technology, China\\
xunyang@nus.edu.sg; yunshan.ma@u.nus.edu; \{liaolizi.llz, eric.mengwang\}@gmail.com; chuats@comp.nus.edu.sg
}

\maketitle
\begin{abstract}
Identifying \textit{mix-and-match} relationships between fashion items is an urgent task in a fashion e-commerce recommender system. It will significantly enhance user experience and satisfaction. However, due to the challenges of inferring the rich yet complicated set of compatibility patterns in a large e-commerce corpus of fashion items, this task is still underexplored. Inspired by the recent advances in multi-relational knowledge representation learning and deep neural networks, this paper proposes a novel Translation-based Neural Fashion Compatibility Modeling (TransNFCM) framework, which jointly optimizes fashion item embeddings and category-specific complementary relations in a unified space via an end-to-end learning manner. TransNFCM places items in a unified embedding space where a category-specific relation (\textit{category}-\textit{comp}-\textit{category}) is modeled as a vector translation operating on the embeddings of compatible items from the corresponding categories. By this way, we not only capture the specific notion of compatibility conditioned on a specific pair of complementary categories, but also preserve the global notion of compatibility. 
We also design a deep fashion item encoder which exploits the complementary characteristic of visual and textual features to represent the fashion products. To the best of our knowledge, this is the first work that uses category-specific complementary relations to model the category-aware compatibility between items in a translation-based embedding space. Extensive experiments demonstrate the effectiveness of TransNFCM over the state-of-the-arts on two real-world datasets. 
\end{abstract}

\begin{figure*}[t]
	\centering
	\includegraphics[width=6.1in]{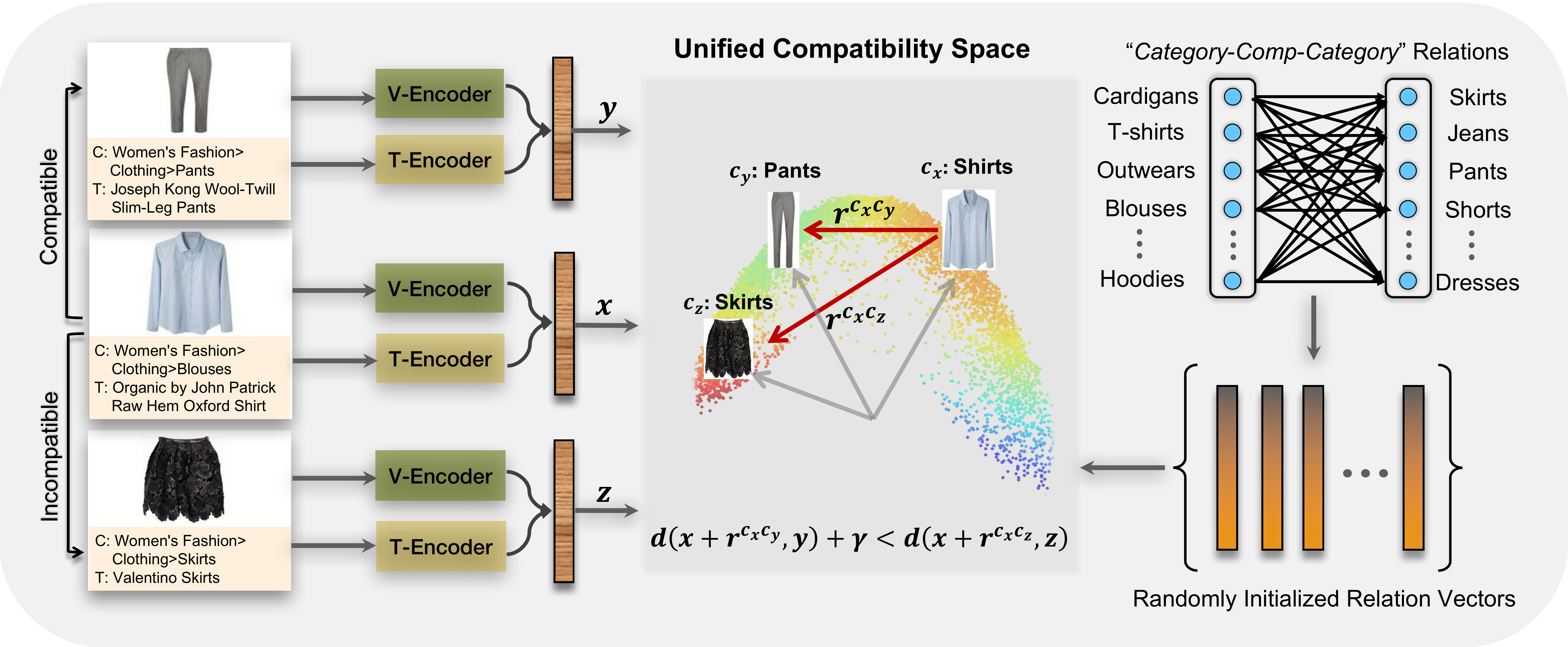}
	\caption{Overview of the proposed Translation-Based Neural Fashion Compatibility Modeling (TransNFCM) approach. It mainly consists of three parts: (1) Each item is first mapped into a latent space by a multimodal item encoder which consists of a pretrained deep CNN for visual modality and a text CNN for textual modality, (2) Category complementary relations (\textit{category-comp-category}) are encoded into the latent space as vector translations operating on the embeddings of compatible items, and (3) Both item embeddings and relation vectors are jointly optimized by minimizing a margin-based ranking criterion.
	} 
	\label{fig2}
\end{figure*} 
\section{Introduction}
Recently, rising demands for fashion products have driven researchers in e-commerce to develop various techniques to effectively recommend fashion items. Existing techniques can be mainly categorized into two types: (1) \textit{search}-based: when a user views a fashion item online, system suggests similar items which the user may also like, and (2) \textit{mix-and-match}-based: when a user views a fashion item (e.g., blouse), the system suggests compatible items (e.g., pants) from a complementary category. 
The first one has already been used in most fashion e-commerce sites, by modeling the visual similarities or interaction relationships such as \textit{also-viewed}. The second approach is more challenging and still underexplored, since it needs to infer the \textit{compatibility} relationships among fashion items that go beyond merely learning visual similarities.
This paper mainly focuses on the second type, termed as cross-category fashion recommendation (CCFR).
It has received increasing attention from multiple research fields, due to its potential ability to enhance user online shopping experience and satisfaction \cite{hsiao2018creating,han2017learning,vasileva2018learning,veit2015learning}.

The key to designing a CCFR model is 1) how to represent various fashion items, and 2) how to model their compatibility relationships based on their representations. Mainstream studies handle this task in an embedding learning strategy: one optimizes a feature mapping from original item space into a latent compatibility space, where compatible items are close together by requiring the pairwise Euclidean distance (inner-product) between embeddings of compatible fashion items to be much smaller (larger) than that of incompatible items \cite{veit2015learning,mcauley2015image,song2017neurostylist,Song2018NeuralSIGIR,he2016learning}. Here, fashion items are represented as a latent vector (also termed as \textit{embedding}) in a latent space, where cross-category compatibility is modeled with pairwise Euclidean distance or inner-product between the embeddings of fashion items. 

Despite their promising performance, such similarity/ metric learning-based compatibility modeling approaches usually suffer from the following limitations. (1) They just consider compatibility learning as a \textit{single-relational} data modeling problem and use a fixed and data-independent function, i.e., Euclidean distance or inner-product, to model the notion of compatibility, which ignores the rich yet complicated patterns in fashion compatibility space. (2) They only utilize pairwise labels (i.e., compatible/incompatible) to optimize item embeddings, thus resulting in a context-unaware compatibility space which ignores the inter-category variation. In such a category-unaware space, incompatible items may be forced to close together in a \textit{one-to-many} case due to the \textit{improper similarity transitivity} \cite{vasileva2018learning}. 
We argue that fashion compatibility is a multi-dimensional concept that varies from case to case. Fashion experts usually use different attributes of items to make a decision for different categories. Data-independent compatibility function usually results in sub-optimal performance.

To overcome these limitations, this paper proposes to design a data-dependent compatibility function which takes category labels of items into modeling. 
Specifically, we formulate compatibility learning as a \textit{multi-relational} data modeling problem in a unified compatibility space dominated by a group of category complementary relations (\textit{category-comp-category}) as shown in Figure \ref{fig2}. 
Here, each relation $\mathbf{r}$ corresponds to a complementary category pair $(c_x, c_y)$ (e.g., T-shirts and Pants), connecting a head item $x$ from $c_x$ to a tail item $y$ from $c_y$. Then, inspired by the highly-celebrated Translation Embedding (TransE) method \cite{bordes2013translating}, we develop a Translation-Based Neural Fashion Compatibility Modeling (TransNFCM) approach by interpreting each relation as a \textit{semantic translation} operating on the low-dimensional embeddings of fashion items. 
Given a pair of compatible items $(x, y)$, our basic learning objective is that the embedding of tail item $\mathbf{x}$ should be close to that of head item $\mathbf{y}$ plus the relation vector $\mathbf{r}^{c_x{c_y}}$, i.e., $\mathbf{x}+\mathbf{r}^{c_x{c_y}}\approx \mathbf{y}$. Finally, we model the category-aware compatibility of $(x, y)$ with a data-dependent distance function $P\big(\!\left(x,y\right)\!\in\!\mathcal{P}\big)\!\propto \! -d\left(\mathbf{x}+\mathbf{r}^{c_x{c_y}},\mathbf{y}\right)$,  which respects the category labels of the head item and tail item and yields a specific compatibility conditioned on $\mathbf{r}^{c_x{c_y}}$. 
Besides, we design a neural item encoder to encode the visual and textual features of an item into a unified embedding vector that are jointly optimized with relation vectors in a latent space. By the proposed TransNFCM, we expect to learn a category-aware notion of compatibility which can better capture complicated compatibility patterns, e.g., \textit{one-to-many}, in a single space.

Our contributions are summarized as follows.
\begin{itemize}
\item We present a novel end-to-end neural network architecture TransNFCM for fashion compatibility modeling. This is the first work that uses category complementary relations to model category-respected compatibility between items in a translation-based embedding space. 

\item We exploit the multimodal complementary characteristic of fashion items with a neural item feature encoder.
\item We conduct extensive experiments on two public datasets, which demonstrates the effectiveness of TransNFCM.
\end{itemize}

\section{Related Work}

\noindent{\textbf{Fashion Compatibility}}. 
In recent years, substantial prior work has been devoted to model the human notion of fashion compatibility for fashion recommendation. Hu et al. (\citeyear{hu2015collaborative}) proposed a personalized outfit recommendation method which models the user-items interaction based on tensor decomposition. A recurrent neural network method in \cite{han2017learning} models outfit composition as sequential process, implicitly learning compatibility via a transition function. Hsiao et al. (\citeyear{hsiao2018creating}) proposed to create capsule wardrobes from fashion images. The above-mentioned approaches focus on outfit compatibility and do not explicitly model item-to-item compatibility.

Mcauley et al. (\citeyear{mcauley2015image}) and Veit et al. (\citeyear{veit2015learning}) proposed to model the visual complementary relations between items by posing it as a metric learning task \cite{yang2018personTIP} that combines Siamese CNNs with co-purchase data in \textit{Amazon} as supervision. Chen et al. (\citeyear{Chen2018AAAI}) proposed a triplet loss-based metric learning method \cite{yang2018person} for fashion collocation. Song et al. (\citeyear{song2017neurostylist,Song2018NeuralSIGIR}) proposed to model compatibility as inner-product between items under Bayesian Personalized Ranking (BPR) framework \cite{he2016vbpr} using co-occurrence relationships in \textit{Polyvore} outfits data as supervision. Both visual and textual modalities are used to represent items with an inter-modality consistency constraint in deep embedding learning manner. In \cite{Song2018NeuralSIGIR}, domain knowledge rules are utilized for compatibility modeling in a teacher-to-student learning scheme \cite{hu2016harnessing}. The main limitation is that they use a fixed and data-independent interaction function to model pairwise compatibility, which ignores the rich yet complicated compatibility patterns and then results in sub-optimal compatibility modeling performance. They will be hard to handle the \textit{one-to-many} case in fashion domain, e.g., they may push incompatible items together improperly when they are compatible with the same target.
\cite{he2016learning} proposed to alleviate this limitation by modeling the notion of compatibility with a weighted sum of pairwise distances in multiple latent metric spaces. 

Different with the aforementioned methods, our work proposes to model category-aware compatibility by jointly optimizing category complementary (co-occurrence) relations and item embeddings in a unified latent space and interpreting the relations as vector \textit{translations} among compatible items from the corresponding categories. The closet work to ours is \cite{vasileva2018learning} that jointly learns type-respected visual item embedding with type-specific sparse masks for modeling both pairwise similarity and category-specific compatibility, in which its category-specific mask projects item pairs into a type-specific subspace for computing compatibility. Our TransNFCM follows a similar motivation with \cite{vasileva2018learning}, but designed in a different way. As analyzed in latter section, TransNFCM not only captures category-specific notion of compatibility but also preserves the global notion of compatibility, thus showing better performance than \cite{vasileva2018learning}.

\noindent{\textbf{Personalized Recommendation}. Our work is also related to personalized (user-item) recommendation methods \cite{he2017neural,he2018nais,he2018outer,tay2018latent} which focus on learning data-dependent interaction functions instead of using a fixed function and show state-of-the-art performance. Our work can also be considered as a content-based recommendation method \cite{chen2017attentive,yu2018aesthetic} where the item features have richer semantics than just a user/item ID. In this paper, we mainly focus on cross-category recommendation in fashion domain. In the future, we will consider to utilize the \textit{user} information (age, shape, weight, religion, etc) for personalized fashion recommendation.

\noindent{\textbf{Knowledge Embedding}. Our method is inspired by recent advances in knowledge representation learning \cite{bordes2013translating,xie2016representation,xie2016representationHie,lin2015learning,nie2015bridging}, where the objective is to model multiple complex relations between pairs of various entities. One highly-celebrated technique TransE \cite{bordes2013translating} embeds relations of multi-relational data as translations operating on the low-dimensional embeddings of entities. TransE has been employed for visual relation modeling \cite{zhang2017visual} and recommendation (Tay et al. \citeyear{tay2018latent}), \cite{he2017translation}.
	Our work is motivated by those findings. We treat each fashion item as an entity in knowledge graphs. Our key idea is to represent the category-aware compatibility relations between items as translations in a unified embedding space, as shown in Figure \ref{fig2}. 


\section{Proposed Approach: TransNFCM}
This paper aims to tackle the task of fashion compatibility learning, which needs to address two sub-problems: 
\begin{itemize}
\item How to effectively represent fashion items that are usually described by multimodal data (e.g., video, image, title, description, etc)?
\item How to effectively model the notion of category-aware compatibility between fashion items?
\end{itemize}
We address these two problems by developing an end-to-end deep joint embedding learning framework, termed as Translation-based Neural Fashion Compatibility Modeling. The overall framework is illustrated in Figure \ref{fig2}. 
Fashion items with multimodal descriptions are encoded as low-dimensional embeddings in a latent compatibility space via a multimodal item encoder. Then, the complicated compatibility relations are captured by a category-specific vector translation operation in a latent space. 
\subsection{Multimodal Item Encoder}
 In the mainstream recommender systems \cite{he2017neural}, only the item ID is used to represent the item, resulting in a high-dimensional and very sparse feature input. While, content-based item representation \cite{chen2017attentive} is more popular in the fashion domain  \cite{he2016vbpr,he2016learning,liao2018knowledge}, since it can capture rich semantical features of fashion items. Generally, fashion items in e-commerce sites are described by multimodal data. This work aims to exploit the complementary characteristic of multimodal descriptions for robust item representation by designing a two-channel item encoder consisting of a visual encoder (\textbf{V-Encoder}), and a textual encoder (\textbf{T-Encoder}). We propose to simultaneously learn two nonlinear feature transformations: one is V-Encoder $f_{V}(v_x)$ that maps an image $v_x$ of item $x$ into a visual feature space $\mathcal{R}^d$ and the other is T-Encoder $f_{T}(t_x)$ that transforms an textual description $t_x$ of item $x$ into a textual feature space $\mathcal{R}^{d}$. We implement both V-Encoder and T-Encoder using the popular deep convolutional network models. Both the outputs of V-Encoder and T-Encoder are $\ell_2$-normalized and concatenated as the final representation $\mathcal{R}^{2d}$ of item $x$ in a feature-level fusion manner.

\subsection{Fashion Compatibility Modeling}	
How to model the item-to-item compatibility is the key of this task. Let's first briefly recall the strategies in prior work. Given a pair of compatible items $(x,y)$ from complementary categories and their feature vectors $\mathbf{x}\in\mathcal{R}^D$, $\mathbf{y}\in\mathcal{R}^D$, the notion of compatibility is modeled as four ways:
\begin{itemize}
	\item \textbf{Inner-product} has been used in previous works \cite{song2017neurostylist,Song2018NeuralSIGIR,he2016vbpr} which model the compatibility as
\begin{equation}\label{eq1}
	P\big(\!\left(x,y\right)\in\mathcal{P}\big)\propto \mathbf{x}^T\mathbf{y},
\end{equation}
where $P\big(\!\left(x,y\right)\in\mathcal{P}\big)$ denotes the probability of $x$ and $y$ being compatible with each other.
	\item \textbf{Euclidean distance} is adopted in \cite{mcauley2015image,veit2015learning,Chen2018AAAI}, which models the compatibility as $P\big(\!\left(x,y\right)\in\mathcal{P}\big)\propto -d(x,y)$, where 
\begin{equation}\label{eq2}
          d(x,y)=\Vert \mathbf{x}-\mathbf{y} \Vert ^2_2=\Vert \mathbf{x} \Vert ^2_2 + \Vert \mathbf{y} \Vert ^2_2 -2\mathbf{x}^T\mathbf{y}.
\end{equation}
    \item \textbf{Probabilistic mixtures of multiple distances} is proposed in \cite{he2016learning} which models the compatibility with a weighted sum of  distances $d_k(x,y)$ in $M$ metric subspaces, parameterized by $M$ matrices $\{\mathbf{E}_k\}_k^M$
 \begin{align}\label{eq3}
 \begin{aligned}
 d(x,y)&=\sum\nolimits_{k} P\left(k|x,y\right)d_k(x,y)\\ 
           &= \sum\nolimits_{k} P(k|x,y) \left\Vert \mathbf{E}_0^T\mathbf{x}-\mathbf{E}_k^T\mathbf{y} \right\Vert ^2_2
 \end{aligned},
 \end{align}
 where  $P(k|x,y)$ denotes the probability of $\mathbf{E}_k$ being used for pair $(x,y)$.
 \item \textbf{Conditional similarity} (Veit et al. \citeyear{veit2017conditional}) is used in \cite{vasileva2018learning} to model the type-aware compatibility with  the  pairwise distance in a conditioned similarity subspace 
 \begin{equation}\label{eq4}
 d(x,y)=\big\Vert \mathbf{x}\odot \mathbf{w}^{c_x{c_y}} -\mathbf{y}\odot \mathbf{w}^{c_x{c_y}}\big\Vert ^2_2,
 \end{equation}
 where $\odot$ denotes element-wise multiplication, $c_i,{c_j}$ denote the types (categories) of item $x$ and $y$, and $\mathbf{w}^{c_i{c_j}}$ is a sparse vector, acting as a gating function that selects the relevant dimensions of the embedding for determining compatibility in the similarity subspace depending on the item type pair $(c_i,{c_j})$. 
\end{itemize}
\noindent{\textbf{Remarks}: Both the first and second types of methods use a fixed and data-independent compatibility function (Eq. (\ref{eq1}) or (\ref{eq2})). When embeddings of item are normalized to unity norm, the compatibility only depends on $\mathbf{x}^T\mathbf{y}$. {It can only capture the \textit{global} notion of compatibility in a shared embedding space, which ignores the rich yet complicated matching patterns in original item space}.  The third measurement (Eq. (\ref{eq3})) adopts a data-dependent function to model the \textit{global} notion of compatibility as a probabilistic mixture of multiple {local} compatibility scores. However, it needs to optimize $M$+2 projection matrices with limited constraints, which can easily get stuck in bad local optima.
Eq. (\ref{eq4}) models the conditioned compatibility by employing a mask operation to embed item pairs into type-specific subspaces for computing conditional similarity, which only captures the \textit{specific} notion of compatibility.

\medskip
\noindent{\textbf{Translation-based Compatibility Modeling}. Most prior work just treats compatibility learning as a problem of single-relational data modeling towards learning a global notion of compatibility. However, fashion compatibility is a multi-dimensional concept that varies from case to case. For example, fashion items belong to various categories (e.g., dresses, coats, skirts, sandals, etc). Given a pair of item from category $A$ and $B$, the visual/semantic attributes which fashion experts use for making a decision may be color, pattern, sleeve-length, material, etc. While, given an item pair from category $C$ and $D$, the attributes they use would change accordingly. In this case, the category labels of items would influence the decision making of experts significantly.
	
To overcome the limitations of existing work, we propose to incorporate category complementary (i.e., co-occurrence) relationships into compatibility modeling, which formulates this task as a problem of multi-relational data modeling, towards learning a category-aware compatibility notion. For simplicity, we term category complementary relationships as \textit{category-comp-category}, such as dresses-comp-boots, which are illustrated in Figure \ref{fig1}. The next question then is how to explicitly represent such relations.
Inspired by the highly-celebrated TransE method \cite{bordes2013translating}, we interpret the \textit{category-comp-category} relations as a simple vector translation operating on the embeddings of compatible items from the corresponding categories. Given a pair of compatible items $(x,y)\in\mathcal{P}$ with embedding vectors $(\mathbf{x},\mathbf{y})$, and the corresponding \textit{category-comp-category} relation vector $\mathbf{r}^{c_xc_y}$, we assume that the embedding of tail item $y$ should be close to that of head item $x$ plus the relation $\mathbf{r}^{c_x{c_y}}$, i.e., $\mathbf{x}+\mathbf{r}^{c_x{c_y}}\approx \mathbf{y}$. Then, our notion of compatibility is modeled with a data-dependent distance function conditioned on $\mathbf{r}^{c_x{c_y}}$, i.e., $P\big(\!\left(x,y\right)\in\mathcal{P}\big)\propto -d\left(\mathbf{x}+\mathbf{r}^{c_x{c_y}},\mathbf{y}\right)$, where
 \begin{align}\label{eq5}
\begin{aligned}
d\left(\mathbf{x}+\mathbf{r}^{c_x{c_y}},\mathbf{y}\right)=&\big\Vert \mathbf{x}+\mathbf{r}^{c_x{c_y}}-\mathbf{y}\big\Vert^2_2\\
=&\Vert \mathbf{x} \Vert ^2_2 +\Vert \mathbf{y} \Vert ^2_2+ \big\Vert \mathbf{r}^{c_x{c_y}} \big\Vert ^2_2\\ &-2\underbrace{{\mathbf{x}^T\mathbf{y}}}_\textrm{global}-2\underbrace{(\mathbf{y}-\mathbf{x})^T{\mathbf{r}^{c_x{c_y}}}}_\textrm{category-specific}
 \end{aligned},
\end{align}
During the training stage, the embeddings of items and relations are jointly optimized in an end-to-end manner.
During the recommendation stage, candidates that have smaller distance (computed using Eq. (\ref{eq5})) with the query item would be top-ranked and suggested to user.}

\noindent{\textbf{Remarks}: Note that, as shown in Eq. (\ref{eq5}), our compatibility modeling function \textit{not only preserves the \textbf{global} notion of compatibility} by $\mathbf{x}^T\mathbf{y}$, \textit{but also captures the \textbf{category-specific} compatibility}  $(\mathbf{y}-\mathbf{x})^T{\mathbf{r}^{c_x{c_y}}}$, which is significantly different with \cite{vasileva2018learning}. Here, the relation $\mathbf{r}^{c_x{c_y}}$ 
functions as a \textit{mask} implicitly that can select category-aware features of items for computing compatibility in a subspace conditioned on the pairwise categories $(c_x, c_y)$.
	
\noindent{\textbf{Margin-based Ranking Criterion}. Given a positive triplet $({x},{y},{r}^{c_x{c_y}})$ consists of two items $(x,y)\in\mathcal{P}$ from category $c_x$ and $c_y$, we generate a set of negative (incompatible) triplets with either the head or tail item replaced by a random item (but not both at the same time). In this way, we can generate a large set $\mathcal{T}$ of 5-tuples for training: 
\begin{align}\label{eq6}
\begin{aligned}
\mathcal{T}=&\big\{({x},{y},{r}^{c_{x}{c_y}}, y', {r}^{c_{x}{c_y'}})|(x,y')\notin{\mathcal{P}}\big\}\cup\\
&\big\{({x},{y},{r}^{c_{x}{c_y}}, x', {r}^{c_{x'}{c_y}})|(x',y)\notin{\mathcal{P}}\big\}
\end{aligned}
\end{align}
where $x'(x)$ and $y(y')$ are not compatible with each other but from complementary categories. To jointly optimize item embeddings and category-specific relation vectors, we minimize a margin-based ranking criterion over the training set
\begin{align}\label{eq7}
\!\!\mathcal{L} = \sum\limits_{\mathcal{T}}\big[d(\mathbf{x}+\mathbf{r}^{c_x c_y},\mathbf{y})-d(\mathbf{x}'+\mathbf{r}^{c_{x'}c_{y'}},\mathbf{y}') +\gamma \big]_+
\end{align}
where $(x',y')\in\big\{(x,y')\cup{(x',y)}\big\}$, and $[\ \cdot\ ]_+$ denotes hinge loss,  and $\gamma>0$ is a margin parameter, and $|\mathcal{T}|$ denotes the total number of 5-tuples in training set. The optimization goal is  that distance between a pair of compatible items should be smaller than that between incompatible (less compatible) items by a margin. 

\noindent{\textbf{Implementation Details}.  The V-Encoder is implemented by the pretrained AlexNet (Krizhevsky et al. \citeyear{krizhevsky2012imagenet}) which consists of 5 convolutional layers and 3 fully-connected (FC) layers. We drop the last FC layer and add a new FC layer as our visual embedding layer that transforms the 4096-D output of the 2nd FC layer into a $d$-dimensional embedding. The T-Encoder is implemented with the text CNN architecture \cite{kim2014convolutional} consisting of a convolutional layer, a max-pooling layer, and a FC layer.  We use four filter windows with sizes of \{2,3,4,5\} with 50 feature maps each.  The textual data are first preprocessed by filtering words appearing in very less items and with very less characters, and then we represent each word with the publicly-available 300-D \textit{word2vec} vector. We use a new FC layer to replace the last layer to transform the output of max-pooling layer into a $d$-dimensional textual embedding ($d=128$). In the multimodal fusion setting, the outputs of the V-Encoder and T-Encoder are $\ell_2$ normalized and then concatenated as the final item embedding $\mathcal{R}^{2d}$. Note that other multimodal feature fusion strategies, such as score-level fusion, can also be used in TransNFCM.
	
We optimize our proposed TransNFCM with the objective Eq. (\ref{eq7}) by stochastic gradient decent (SGD) in minibatch training mode. All the visual/textual item embeddings are normalized to unit norm. For each pair of complementary categories (i.e., t-shirts and pants), we generate a category-aware relation vector, having the same dimension with item embeddings. The relation vectors are randomly initialized and only normalized to unit norm at the beginning of optimization. No other regularization or norm constraints are enforced on relation vectors. TransNFCM is implemented with Pytorch. In each epoch, we first shuffle all the 5-tuples in $\mathcal{T}$ and get a mini-batch in a sequential way.

\begin{table*}[htbp]\small
	\caption{Comparison on the FashionVC and PolyvoreMaryland datasets based on two metrics: AUC (\%) and Hit@K (\%, K$\in$\{5, 10, 20, 40\}). A larger number indicates a better result.  \textbf{V} and \textbf{T} denote \textbf{V}isual modality and \textbf{T}extual modality, respectively. \textbf{V+T} denotes the fusion of visual and textual modalities. \textit{100 negative candidates are sampled for each query during testing.} The best results are shown in boldface.
	}
	\begin{center}	
		{\begin{tabular}{c|c|c| c c c c|c| c c c c} 
		        \hline		
				\multirow{2}{*}{{Features}} & \multirow{2}{*}{{Methods}}   &\multicolumn{5}{c|}{{FashionVC}} &\multicolumn{5}{c}{ {PolyvoreMaryland}}  \\\cline{3-12}
				&& AUC & Hit@5 & Hit@10 & Hit@20 & Hit@40 & AUC & Hit@5 & Hit@10  &Hit@20 & Hit@40\\
			   \hline
				\multirow{5}{*}{V} &{{SiaNet}}     &60.4&9.7&18.1&31.2  &52.8  &59.1     &8.3     &15.5         &29.0     &51.8  \\
				&{{Monomer}}     &70.2     &16.9   &28.6   &45.8    &69.1  & 70.5     & 17.6       &28.9        &45.7          &69.0    \\
				&{{CSN}}   & 71.6    &  16.7   &  28.4   & 46.7    &  70.8  &70.2       &17.3     & 28.4   & 45.1      &68.4   \\
				&{{BPR}}   &70.9  &16.7  &27.3     &46.7          &70.4   &69.5    & 17.3    &28.2          &43.9              &67.5   \\
					&{{TriNet}}     &70.6    &16.3       &28.0         &45.7                &69.6     &70.1 & 18.1   &28.7            & 44.9           &68.3   \\
				&{{\textbf{TransNFCM}}}   &{73.6}     &  {19.0 }  &  {32.3}   &  {51.6}   &  {74.0 } &{71.8}       &  {18.9}    & {30.6} &{48.1}    & {70.5} \\
				\hline
				\hline
				\multirow{5}{*} {T}   &{{SiaNet}}    &  66.1   &  10.8   &  21.0   &  37.9   &  61.1  &62.3   &8.3   & 16.2   & 32.0      &56.3  \\
				&{{Monomer}}     &68.8     &16.5     & 26.9    &  42.1   & 64.8   &63.3    &10.1    &18.8   & 33.9     & 58.1  \\
				&{{CSN}}  & 67.5    &11.2     &22.4     & 41.2    & 64.1   &63.2      &8.8    &17.0   &32.5       &57.4   \\
				&{{BPR}} &  70.9   & {15.4}    &   26.8  &   45.6  &  67.6  &  67.8   &  13.0  &  23.6 &  40.3    & 65.3 \\
				&{{TriNet}}    & 71.3   &  16.5   &  28.9   &   46.4  & 69.2   &   68.4   &  13.7  &  24.4 &  41.5    & 65.8 \\
				&{{\textbf{TransNFCM}}}   &   {72.6}  &  {18.9}   &  {30.0}   &  {47.9}   & { 70.8} &   {68.8}   &   {14.7}  & {25.8}  &  {42.2}   &  {66.0}\\
				\hline
				\hline
				\multirow{1}{*} {V+T}  	
				&{{\textbf{TransNFCM}}}   &  {76.9}   &  {23.3}   & {38.1}    &  {57.1}   & 77.9  &74.7      &21.7     &34.4   &52.7     &75.3   \\
				\hline                                                     	
		\end{tabular}}
	\end{center}
	\label{Table1}
\end{table*}

\begin{figure}[t]
\centering
\includegraphics[width=3.3in]{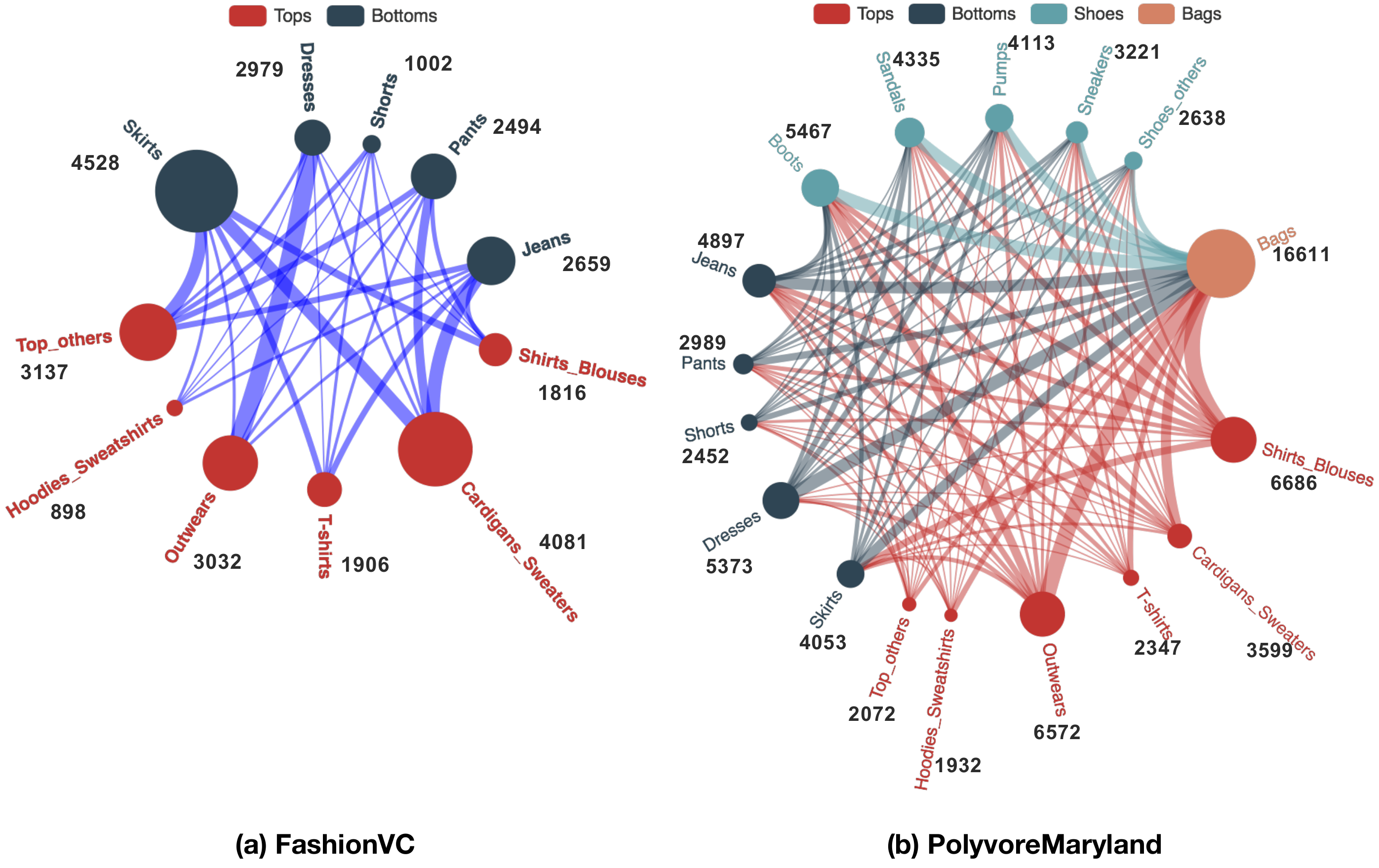}
\caption{Illustration of \textit{category}-\textit{comp}-\textit{category} relations in both datasets: (a) FashionVC, (b) PolyvoreMaryland. Each edge denotes a \textit{category-comp-category} relation. The numbers of items in each category are also illustrated. For simplicity, \textit{Dresses} is classified into the \textit{Bottoms}. 30 \textit{category}-\textit{comp}-\textit{category} relations are used in FashionVC and 101 relations are used in PolyvoreMaryland.} 
\label{fig1}
\end{figure}
\section{Experiments}
To comprehensively evaluate the effectiveness of our proposed TransNFCM method, we conduct experiments to answer the following research questions:\\
\noindent{\textbf{RQ1}: Can our proposed TransNFCM outperform the state- of-the-art fashion compatibility learning methods? \\
\noindent{\textbf{RQ2}: Is the multimodal embedding fusion strategy helpful for improving the learning performance? \\
\noindent{\textbf{RQ3}: Why does TransNFCM work?

\subsection{Experimental Settings}
\noindent{\textbf{Datasets}}. We conduct experiments on two public fashion compatibility datasets, both crawled from the fashion social commerce site \textit{Polyvore} (www.polyvore.com), which enables fashion bloggers to share their fashion tips by creating and uploading outfit compositions. In the following two datasets, we both use the item pairs that are co-occurring in an outfit as supervision for training and also evaluation.\\
\noindent{\textbf{FashionVC}}} \cite{song2017neurostylist}. It was collected for top-bottom recommendation, consisting of 14,871 tops and 13,663 bottoms, split into 80\% for training, 10\% for validation, and 10\% for testing. Each item contains a product image and textual description (category and title).\\
\noindent{\textbf{PolyvoreMaryland}} \cite{han2017learning}. It was released by \cite{han2017learning} for outfit compatibility modeling. Since in this paper we mainly study the modeling of item-item compatibility, not the whole outfit, we only keep four groups of items among the first 5 item in outfits: tops, bottoms, shoes, and bags. Other fashion accessories and jeweleries have been removed. Each item contains a product image and textual description (product title only). We extract and re-split the co-occurring item pairs randomly in the same setting with FashionVC: 80\% for training, 10\% for testing, and 10\% for validation.

\noindent{\textbf{Evaluation Protocols}}. For each testing positive pair $(h_i,t_{ig})\in{\mathcal{P}_t}$, we replace the tail item with $N=100$ negative items\footnote{Note that our setting is much more challenging that that of \cite{song2017neurostylist,Song2018NeuralSIGIR} where only 3 negative candidates are sampled for each query during testing.} $\{t_{in}\}^N_{n=1}$  which do not co-occur with $h_i$ in the same outfit but are from complementary categories with $h_i$. We adopt two popular metrics, \textit{Area Under the ROC curve} (AUC) and \textit{Hit Ratio (HR)}, to evaluate the item-item recommendation based on the compatibility score. Both AUC and HR@$K$ are widely-used in recommendation systems. AUC is defined as
\begin{equation}
\textrm{AUC}=\frac{1}{N|\mathcal{P}_t|}\sum\limits_{i}\sum\limits_{n}\delta\big(s(h_i,t_{ig})>s(h_i,t_{in})\big)
\end{equation}
where $\delta(a)$ is an indicator function that returns 1 if the argument $a$ is true, otherwise 0. $s(h_i,t_{ig})=P\big(\!\left(h_i,t_{ig}\right)\in\mathcal{P}_t\big)$ denotes the compatibility score. $|\mathcal{P}_t|$ denotes the total number of testing positive pairs. HR@$K$ is a recall-based metric, denoting the proportion of the correct tail item $t_{ig}$ ranked in the top $K$. $N+1$ candidates are provided for each head item $h_i$.

\noindent{\textbf{Parameter Settings}}. We employ the stochastic gradient descent for optimization with momentum factor as 0.9. We set the overall learning rate $\eta=0.001$, and drop it to $\eta=\eta/10$ every 10 epochs. 
The learning rate of the pretrained AlexNet in V-Encoder is set to $\eta'=\eta/10$ for fine-tuning. The margin $\gamma$ is set to 1 following the setting of \cite{bordes2013translating}.
We use 128 5-tuples in a minibatch. Both the dimensions of visual embedding vectors and textual embedding vectors are set to $d=128$.  Dropout is used in both visual and textual encoders.

\noindent{\textbf{Comparison Methods}}. We compare TransNFCM with the following methods that are all implemented in the same framework. The main difference lies in the compatibility modeling functions and loss functions. 
\begin{itemize}
\item[-] {\textbf{Siamese Network} (SiaNet)} \cite{veit2015learning}: models compatibility with the squared Euclidean distance (Eq. (\ref{eq2})) and uses contrastive loss as its optimization criterion.
\item[-] {\textbf{Triplet Network} (TriNet)} \cite{wang2014learning,Chen2018AAAI}: models compatibility with Eq. (\ref{eq2}) and use the margin-based ranking criterion with $\ell_2$-normalized item embeddings as input, in which the margin is set to 1. TriNet is our baseline which is category-unaware and uses data-independent function. 
\item[-] {\textbf{BPR}} \cite{he2016vbpr,song2017neurostylist}: models compatibility as inner-product (Eq. (\ref{eq1})) and uses the soft-margin based objective loss. In this paper, we implemented it with $\ell_2$-normalized embeddings.
\item[-] {\textbf{Monomer}} \cite{he2016learning}: models compatibility with a mixture of multiple local distances (Eq. (\ref{eq3})) and is also implemented using the same objective function with us.
\item[-] {\textbf{CSN}} \cite{vasileva2018learning}: models pairwise compatibility as a conditional similarity (Eq. (\ref{eq3})) that respects item types. We implement it following the setting of \cite{veit2017conditional}.
\end{itemize}

\subsection{Experimental Evaluation}
 \noindent{\textbf{Performance Comparison and Analysis}}. Table \ref{Table1} shows the performance comparison on two datasets based on AUC and Hit@K (K$\in$\{5,10,20,40\}). 
 From Table \ref{Table1}, we have the following observations. 
 \begin{itemize}
 \item {TransNFCM achieves the best performance} in most cases and obtains high improvements over the comparison methods, especially in the visual (V) modality setting. This justifies the effectiveness of TransNFCM that builds a data-dependent (i.e., category-aware) compatibility function (Eq. (\ref{eq5})) using a translation-based joint embedding learning paradigm. (\textbf{RQ1})
 
\item TransNFCM consistently outperforms the baseline TriNet on two datasets, especially using visual features. It demonstrates the necessity of encoding the pairwise category-labels into the embedding space for capturing a category-aware compatibility notion. Note that the improvement becomes less significant in the T setting on the PolyvoreMaryland dataset. It is mainly because that the textual descriptions of items in the PolyvoreMaryland dataset are very noisy and sparse, thus resulting in sub-optimal relation embeddings. (\textbf{RQ1})

\item Although both employing the pairwise category labels, TransNFCM significantly outperforms CSN in all. It is mainly because CSN only captures the conditioned compatibility in a subspace dominated by a pair of complementary categories and does not preserve the global notion of compatibility in original feature space, thus resulting in a limitation that its conditioned compatibility in one subspace cannot be compared with that in another subspace. While, benefiting from the TransE framework, our compatibility function (Eq. (\ref{eq5})) not only captures the category-specific notion but also preserves the global notion, thus showing better performance. (\textbf{RQ1})

\item Table \ref{Table1} also shows that TransNFCM can effectively mine the complementary characteristic of images and texts with the multimodal item encoder. The textual descriptions of items in the two datasets are very noisy and also sparse. But it indeed contributes some visually-invisible attributes such as \textit{style} or \textit{season} that can well complement visual features, thus bringing significant performance improvement. Note that BPR has been used for fashion compatibility modeling in a multimodal fusion manner in \cite{song2017neurostylist}. We have conducted a performance comparison with BPR in our multimodal fusion setting with 100 negative candidates: BPR achieves 75.4\% AUC and 33.4\% Hit@10 on the FashionVC dataset, and 73.8\% AUC and 32.7\% Hit@10 on the Polyvore dataset. TransNFCM yields remarkable improvements on BPR.
Besides, we do not include a cross-modality consistency loss \cite{song2017neurostylist} or visual-semantic loss \cite{vasileva2018learning} in TransNFCM, since we empirically found that those strategies would reduce the diversity of multimodal item representation and degrade the performance. 
(\textbf{RQ2})
 \end{itemize}

\begin{figure}[t]
	\centering
	\includegraphics[width=3.3in]{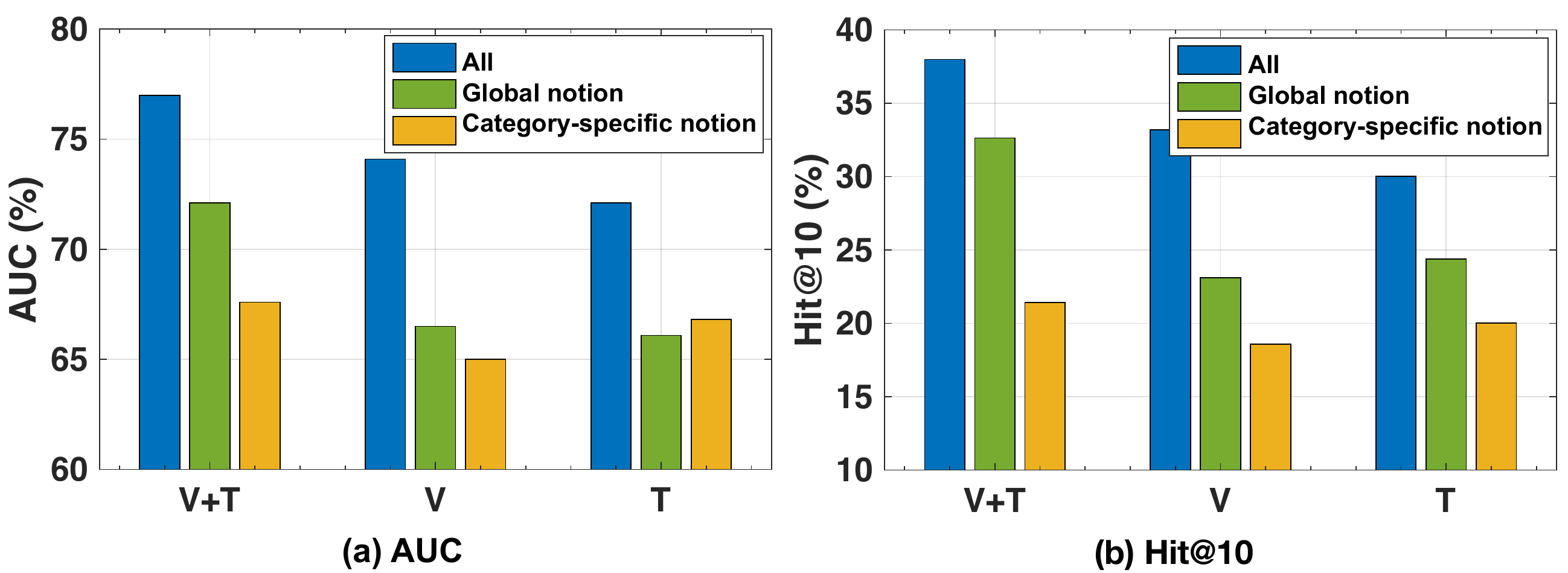}
	\caption{Effects of using different parts in Eq. (\ref{eq5}) for compatibility computing on FashionVC.  \textit{{Global notion}} refers to  using  $\mathbf{x}^T\mathbf{y}$, \textit{{Category-specific notion}} refers to using $(\mathbf{y}-\mathbf{x})^T\mathbf{r}^{c_xc_y}$, and  \textit{{All}} refers to using Eq. (\ref{eq5}). }
	\label{fig3}
\end{figure}

\noindent{\textbf{Empirical analysis.}}  To better evaluate the effects of the global part and category-specific part in Eq. (\ref{eq5}), we conduct an experiment (as shown in Figure \ref{fig3}) to investigate how these jointly-learned two parts performs separately in the testing stage. From Figure \ref{fig3}, we have the following observations. $\mathbf{x}^T\mathbf{y}$ usually performs better than $\mathbf{(y-x)}^T\mathbf{r}^{c_xc_y}$ in all except one, which demonstrates the importance of preserving such global notion of compatibility. The category-specific part complements the global part well, since it alleviates the \textit{improper similarity transitivity} by pushing hard negative candidates farther away from the query. Only using $\mathbf{(y-x)}^T\mathbf{r}^{c_xc_y}$ results in a bad performance, which is also validated by the performance of CSN in Table \ref{Table1}. In TransNFCM, $\mathbf{r}^{c_xc_y}$ dominates a latent compatibility subspace where compatible items $(x,y)$ from categories $(c_x,c_y)$ are close to each other. So, we empirically conclude that it is the combination of both global notion and category-specific notion of compatibility that makes TransNFCM work. (\textbf{RQ3})

\noindent{\textbf{New setting: Target category is known.}} We also compare TransNFCM with TriNet in a new evaluation setting: \textit{the target category is known}, e.g., a user buys/clicks a blouse and wants system only recommend skirts that match the given blouse well. This new setting facilitates our baseline TriNet more, since all the negative candidates are sampled from the known target category. The \textit{improper similarity transitivity} in TriNet can be alleviated. We observe from Figure \ref{fig5} (a) and (b) that our TransNFCM can still outperform TriNet by nearly 2\% AUC and Hit@10. When using the original setting (negative candidates are randomly sampled from complementary categories), shown in Figure \ref{fig5} (c) and (d), the improvement becomes more obvious. Nearly 4\% improvement is observed at both AUC and Hit@10.

\begin{figure}[t]
	\centering
	\includegraphics[width=3.3in]{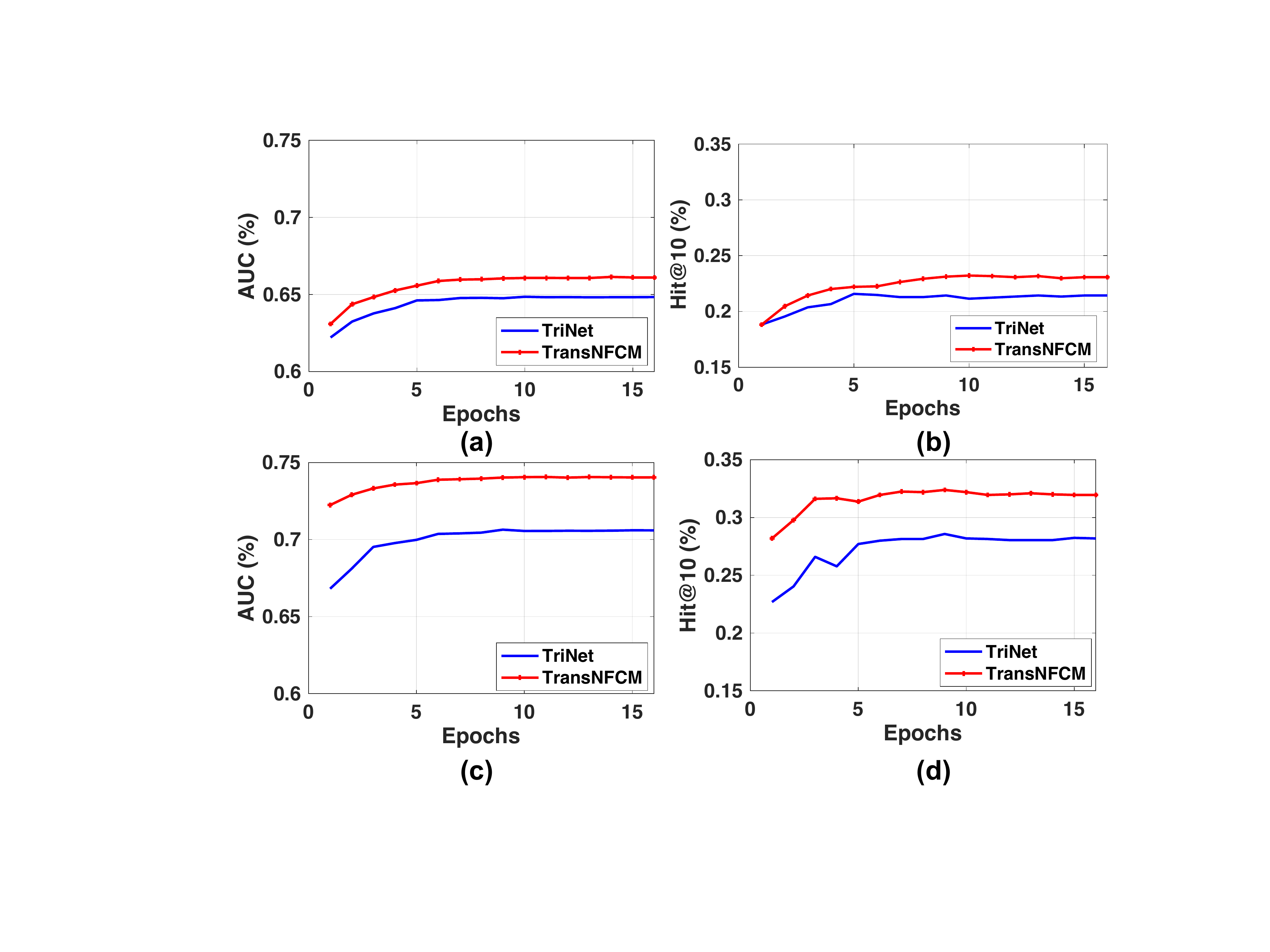}
	\caption{ (a) (b) denote the comparison with TriNet when target category is known, (c) (d) denote the comparison with TriNet when target category is unknown. Experiments are conducted on FashionVC. Only visual modality is used. }
	\label{fig5}
\end{figure}
\section{Conclusion}
In this work, we developed a new neural network framework for fashion compatibility modeling, named TransNFCM. The main contribution is that TransNFCM casts fashion compatibility as a multi-relational data modeling problem and encodes the \textit{category-comp-category} relations as vector translations operating on embeddings of compatible items in a latent space. TransNFCM not only captures the category-specific notion of compatibility but also preserves the global notion, which can effectively alleviate the \textit{improper similarity transitivity} in metric learning based approaches. To the best of our knowledge, this is the first work that poses compatibility modeling into such a translation-based joint embedding learning framework. Although TransNFCM only utilizes category-level co-occurrence relations in this work, it can be directly extended to model fine-grained matching rules, composed of color, category, pattern, style, etc. We will explore more fine-grained relations in the fashion domain to further discover the potentials of TransNFCM in the future.
Besides, we also introduce a multimodal item encoder which effectively exploits the complementary characteristic of different modalities. Extensive experiments on two datasets have demonstrated the effectiveness of our method. 
\section{ Acknowledgments}
We would like to thank all reviewers for their comments. NExT research is supported by the National Research Foundation, Prime Minister's Office, Singapore under its IRC@SG Funding Initiative. This work is also supported by the National Nature Science Foundation of China (NSFC) under grant 61732008 and 61725203.

 \bibliographystyle{./aaai}
\bibliography{IEEEexample}

\end{document}